\documentclass{mn2e}
\usepackage{graphicx}

\title{A Study of Some Current Methods of Analysing Observations of
Star Forming Regions}
\author[S. D. Doty \& M. L. Palotti]
       {S. D. Doty \& M. L. Palotti \\
        Department of Physics and Astronomy, Denison University,
	Granville, OH  43023, USA} 
\date{Accepted 2002 May 13; 
      Received 2002 May 11;
      in original form 2002 February 13}

\pagerange{\pageref{firstpage}--\pageref{lastpage}}
\pubyear{2002}

\begin{document}

\maketitle

\label{firstpage}

\begin{abstract}
We present an evaluative study of some current methods utilized
in the analysis of infrared (IR) observations of star-forming
regions.  A series of self-consistent radiative transfer models
are constructed, with the outputs analysed using these methods
to infer source properties such as dust tempertaure, mass, opacity function,
and density distribution.  Any discrepancies between the inferred
and model quantities can be attributed to the analysis methods.
The range of validity of most methods is 
smaller than expected, due to two effects:  (1) limited applicability 
of the Rayleigh-Jeans limit except to very long wavelengths, and (2) 
significant errors in the isothermal approximation, even when 
$\Delta T(r) < 2\mathrm{K}$ over 90 per cent of a region.  Still, 
an accurate mean $T_{\mathrm{dust}}$ can be found using a 
modified Wiens law.  This temperature can yield dust masses 
to within 10-25 per cent -- much better than masses inferred from 
the integrated luminosity. Using long wavelengths 
($> 1000-2000 \mu \mathrm{m}$),
the opacity index can be determined from the far-IR 
spectrum to within 20 per cent. Fitting the spectrum yields better 
results.  The density
distribution can be somewhat constrained by fitting the surface brightness,
for well-resolved sources. Better
results are found by fitting the flux spectrum with detailed models.

\end{abstract}


\section{Introduction}
Infrared (IR) radiation, due to thermal emission from 
dust grains, is a key feature of star-forming regions.  
These grains play an important role in the chemistry, 
thermal balance, and evolution of these regions.  
This role, coupled with the ability for IR radiation
to escape from areas of high optical depth, makes
the determination of source parameters through the
study of IR observations of great importance.

The problem of obtaining source parameters from 
the IR observations is a difficult one.  As a result,
it is important to study and evaluate the analysis methods
used to judge their value and regimes of validity.

In general, the analysis of IR observations can be 
broadly broken into two groups.  The first approach 
relies upon the use of relatively simple, semi-analytic
expressions which are derived from idealized and simplified
approximations and assumptions (see e.g. Hildebrand 1983).
It has been shown previously that the strict use of this 
approach may lead to interpretations which are ambiguous
or erroneous (see e.g., Schmid-Bergk \& Scholz 1976; 
Butner et al. 1991; 
Doty \& Leung 1994; 
Men'shcikov \& Henning 1997;
Shirley, et al. 2000).  
The second approach involves the construction (usually
a grid) of detailed, self-consistent radiative transfer
models for the computation of model spectra and comparison
with observations.  While such codes are available for use
by the general community (e.g. CSDUST3 by Egan, Leung, 
\& Spagna 1983; DUSTY by Ivezic, Nenkova, \& Elitzur 1999), constructing
such models requires substantial time and effort.  As a result,
the first approach remains a common choice.

In this paper we critically evaluate semi-analytic methods
of analysis of IR observations of star-forming regions.
Here we consider internal and externally heated
dust clouds as analogues of early-type star-forming regions.
In particular, we examine the usual assumptions of 
homogeneities in density and temperature, as well as the
neglect of opacity effects when applied to externally 
heated IR sources.  We do this by constructing a series of
realistic models of these regions, treating the 
model output as simulated observations.  We then apply common
semi-analytic methods to infer the source properties, and
compare them to those used in constructing the models.  
The discrepancies between the input and inferred source
properties yield a measure of the limitations of the 
semi-analytic methods used.

The outline of this paper is as follows.  In Section 2 we
summarize the problem and current methods of analysis.
We describe the models run and testing procedure in Section 3.
In Section 4, we discuss the role and determination of
the dust temperature.  We consider the dust mass in Section 5.
In Section 6 we discuss the determination of the dust 
opacity function.  We consider the determination of the dust
density distribution in Section 7.  
Finally, in Section 8, we summarize
and draw conclusions from this study.

\section{Analytical Description of the Problem and 
         Summary of Some Current Methods of Analysis}

\subsection{Emergent radiation}
For a spherically symmetric dust shell, the specific luminosity, 
defined to be the amount of energy emitted
per unit frequency per unit time can be written as
\begin{equation}
L_\nu = \int{<Q_a \pi a^2>_\nu B_\nu [T(r)] e^{-\tau (\nu ,r)} n(r)4 \pi r^2 
~ \mathrm{d}r}.
\end{equation}
Here $B_\nu [T(r)]$ is the Planck function at the dust temperature, $T(r)$,
and $\tau_\nu$, $<Q_a \pi a^2>_\nu$, and $n(r)$ are the optical depth,
grain absorption coefficient, and density distribution respectively as defined
below.  The density distribution and the grain absorption cross section 
 are both assumed to be power laws with indices $m$ and $\beta$, given by
\begin{equation}
n(r) = n_0(r_0/r)^m, 
\end{equation}
and
\begin{equation}
Q_\nu \equiv <Q_a \pi a^2>_\nu = Q_0(\nu / \nu_0)^\beta,
\end{equation}
respectively.  The optical depth, $\tau_\nu$, at a radius $r$ is defined as
\begin{equation}
\tau_\nu = \int{Q_\nu n(r) dr}.
\end{equation}
For convenience, we define the optical depth in the visible to be $\tau_0 
\equiv \tau_{0.55 \mu \mathrm{m}}$.

In order to evaluate the specific luminosity explicitly, the 
temperature distribution, $T(r)$, density distribution, 
and grain properties need to be known.
For this reason, it is common to assume that sources are 
isothermal and optically thin.
In this limit, equation (1) simplifies to 
\begin{equation}
L_\nu = Q_\nu B_\nu (T) \int {n(r) 4 \pi r^2 dr} = Q_\nu B_\nu(T) 
\frac{M_\mathrm{dust}}{m_\mathrm{grain}},
\end{equation}
and can then be solved analytically if the density distribution is known 
or assumed to be known.  From equation (5)
we can derive semi-analytic expressions for many of the source parameters, as
discussed below.

\subsection{Dust temperature}
As stated above, it is common to assume that sources are isothermal
(see, e.g., Launhardt et al. 1996) 
.  
In order to determine
the temperature of such a source,  
Wiens law denoting the peak of an unmodified blackbody spectrum ($B_{\nu}$)
is often used (assuming an isothermal source as
in eq. [5]),
\begin{equation}
T = \frac{2898}{\lambda_\mathrm{peak}(\mu \mathrm{m})}\mathrm{K}.
\end{equation}
It is also possible to find the temperature by fitting 
the flux spectrum (see Sect 2.4).  
In practice this takes many forms, depending upon the quantity
and quality of data available for a given source.
These approaches range from fitting the
observed flux at a single wavelength point to find a single
equivalent brightness temperature (e.g. Henning et al. 2000),
to using a two-wavelength effective color temperature
(e.g. Grady et al. 2001) 
to fitting the flux spectrum with a single temperature greybody
(e.g., Siebenmorgen, Kr\"ugel, \& Chini 1999), 
to using two temperature components
(e.g., Kr\"ugel et al. 1998; Ward-Thompson et al. 2000).  
In many cases, the dust properties are assumed to be known
{\it a priori} (e.g., Kr\"ugel et al. 1998; Henning et al. 2000),
though they are sometimes determined directly from the observations 
(see e.g., Launhardt, Ward-Thompson, \& Henning 1997; 
Voshchinnikov \& Kr\"ugel 1999; 
Abraham et al. 2000; and Sect 2.4).

\subsection{Dust mass}
Assuming a dust grain mass $m_{\mathrm{grain}}$, 
the total dust mass of the source is
\begin{equation}
M_\mathrm{dust} = m_\mathrm{grain} \int{n(r) 4 \pi r^2 dr}.
\end{equation}
In the optically thin and isothermal limits, the dust mass can be expressed in
terms of the specific luminosity (see eq. [5]) by
\begin{equation}
M_\mathrm{dust} = (m_\mathrm{grain} L_\nu)/ (B_\nu Q_\nu).
\end{equation}
By expressing the specific luminosity in terms of the observed flux, 
$f_{\nu,\mathrm{obs}}$, and 
the distance to the source, $D$, the mass can be written as
\begin{equation}
M_\mathrm{dust} = (m_\mathrm{grain} f_{\nu,\mathrm{obs}} D^2)/(B_\nu Q_\nu)
\end{equation}
(Hildebrand 1983;
 in various forms by, e.g, Beichman et al. 1990,
 Launhardt et al. 1996; Siebenmorgen, Kr\"ugel, \& Chini 1999).
As can be seen, the dust mass can be determined from the observed 
flux and an estimate
of the distance to the source.  Unfortunately, this requires a good 
estimate of 
both the dust temperature (see Sect. 4) and grain opacity function 
(see Sect. 6) for an
accurate dust mass estimate.

The mass of the dust can also be determined by the spectral energy 
distribution (Doty and Leung 1994).
The energy emitted from the dust is simply a reprocessing of the energy 
that the dust
has absorbed.  As a result, in an optically thin source the total
emission is related to the total amount 
absorbed, which varies with optical depth or dust mass.
\begin{equation}
L \equiv \int{L_\nu ~ \mathrm{d}\nu} \propto M,
\end{equation}
assuming that the temperature is not a function of the dust mass.

\subsection{Grain opacity function}
For an optically thin source and when the Rayleigh Jeans approximation 
($h\nu \ll kT$; RJA)
holds, the specific luminosity becomes
\begin{equation}
L_\nu \propto \nu^{2+\beta} \int {T(r) n(r) 4 \pi r^2 dr}.
\end{equation}
By taking the ratio of luminosities at two different frequencies, the opacity index, $\beta$
can be determined (Helou 1989) as
\begin{equation}
\beta = [\mathrm{log}(L_{\nu 1}/ L_{\nu 2})/ \mathrm{log} (\nu_1/\nu_2)] - 2.
\end{equation}

An interesting predictor-corrector-like iteration method was utilized by 
Siebenmorgen, Kr\"ugel, \& Chini (1999).  In this approach,
previous greybody models of similar sources were used to guess a
single dust temperature.  This was used to infer a dust opacity
index, $\beta$.  Given this value of $\beta$ the data were fit to
find $T$.  Finally, the corrected value dust temperature was used
to find a final value for $\beta$.  While each step of the iteration
used methods/assumptions described elsewhere here, the iterative
approach helped to break the degeneracy between temperature and
dust opacity.

On the other hand, given the power of modern computers, and the 
large amounts of high quality far-infrared and submillimeter data becoming
available for many sources, it is possible to consider the coupled problem
and determine
the temperature and the opacity index by fitting 
the observed spectrum
(see, e.g., Krugel et. al. 1998; Ward-Thompson et al. 2000; 
Abraham et al. 2000).
For an isothermal, optically thin source, the specific
luminosity can be written as
\begin{equation}
L_\nu = a_1 \nu^{a_3} B_\nu (a_2),
\end{equation}
where $a_1$, $a_2$, and $a_3$ become fitting parameters.  Here, the density
distribution is incorporated into $a_1$, the temperature is $a_2$, and the 
opacity index, $\beta$, is $a_3$.
To compute values for the fitting parameters, the chi-squared
\begin{equation}
\chi^2 = \sum [y_\mathrm{actual}(i)-y_\mathrm{fit}(i)]^2/\sigma^2(i),
\end{equation}
is minimized for a given set of frequencies.  The uncertainty in 
the measure of 
luminosity ($y(i)$) is defined by the parameter $\sigma (i)$.  We have only 
considered the case where the uncertainty is equal 
to a small fraction of $y(i)$ (Doty and Leung 1994).  This has the
effect of equally weighting each point in the spectrum.

\subsection{Dust density distribution index}
If a resolved source is assumed to be isothermal along any line 
of sight, the mean 
intensity can be written as 
\begin{equation}
I_\nu = \tau_\nu B_\nu .
\end{equation}
Adopting a power law form for the optical depth, 
$\tau_\nu \propto Q_\nu \propto \nu^\beta$, and taking
the ratio at two different frequencies, equation (15) is then
\begin{equation}
\frac{I_{\nu 1}}{I_{\nu 2}} = \frac {\nu_1^\beta}{\nu_2^\beta} 
\frac{B_{\nu 1}}{B_{\nu 2}}
\end{equation}
Assuming that the optical index, $\beta$, is known (see Sect. 6), 
a temperature can be found
given a ratio of intensities.  Using this temperature in equation (15) 
will yield
the optical depth for that frequency and line of sight.  
The optical depth at a given impact parameter
is related to the column density, $N$, along that line of sight.  
Therefore, the form
of the optical depth map will be the same form as a column density map 
for the same source.
Finally, the power law form for the column density, 
$N(r) \propto r^{-\alpha}$, can
be related to the power law form of the density distribution, 
$n(r) \propto r^{-m}$ (Tomita et al. 1979) 
where $m=1+\alpha$.  As Yun and Clemens (1991) point out, however, 
this is only appropriate
for an infinite source.
As one probes closer to the edge of a finite
source, the column density along each line of sight approaches zero.  
Thus, when
fitting these outer radii, $\alpha$ is artificially increased. 
It is possible to relate the theoretical column density to the density 
power index by
\begin{equation}
N(p) \propto \int{(z^2+p^2)^{-m/2}dz},
\end{equation}
where $p$ is an impact parameter and $z$ is the distance along the line 
of sight.  For different
values of the density distribution index, a column denisty map is formed.
Following Yun and Clemens (1991), by
matching the derived column density map with the calculated map, 
a value of $m$ can be found.

\section{The Testing Procedure}
In order to evaluate the reliability of current semianalytic methods
of analysis one must compare the derived source parameters with their true
values.  As the source parameters for actual observations are not
known, we have 
created simulated observations using a modified version of the 1D radiative
transfer code of Egan, Leung $\&$ Spagna (1988).  
The output of these models are 
treated as observational data and analysed using current techniques to derive
the source parameters.  These derived results can then be compared with the 
original model parameters.  Any differences between the two sets of data can
be attributed to the method of analysis.

\subsection{Model parameters}
We model sources in the interstellar medium that are
predominantly heated extenally by the interstellar radiation field.  Regions
of star formation, such as many dark clouds or Bok globules, are the
primary motivation.  The processes involved in low-mass star 
formation are 
relatively well understood (see e.g. review by Shu, Adams, \& Lizano 1987).
On the other hand, while the picture is not as complete in the high-mass case 
(see e.g. Churchwell 1993, 1999), recent work 
(van der Tak et al. 2000; Doty et al. 2002a) seems to suggest that 
in general similar scalings may be at least potentially reasonable.
In both situations, an understanding of the source parameters 
(dust properties,
material distribution, existence/nature of an embedded star) are important
(see e.g., Andre, Ward-Thompson, \& Barsony 2000).
Typical model parameters
 are given in Table 1.

%_________________________BEGIN   ________________One column table
   \begin{table}
      \caption[]{Model Parameters}
         \label{modelparameters}
%     $$ 
%         \begin{array}{lll}
         \begin{tabular}{ll}
            \hline
            Parameter & Value \\ 
            \hline
     Cloud size $[r_\mathrm{out}]$ & 1.0 pc \\
     Cloud thickness $[r_0/r_\mathrm{out}]$ & $10^{-5}$ \\
     Cloud opacity $[\tau_0~ at~ 0.55 \mu \mathrm{m}]$ & 10-300 \\
     Density distribution $[n(r)]$ & $n_0(r_0/r)^m, m$ = 0, 1, 2 \\
     Opacity function & $Q_0(\lambda_0/\lambda)^{\beta}, \beta = 1.5$ \\
     Luminosity of internal source $[L_\ast]$ & 0-300 $L_\odot$ \\  
            \hline
     \end{tabular}\\
%         \end{array}
%     $$ 
%\begin{list}{}{}{}
%\item[$^{\mathrm{ }}$] $a(b)$ means $a \times 10^{b}$
%\item[$^{\mathrm{ }}$] All abundances are relative to H$_{2}$,
%\item[$^{\mathrm{b}}$] This is footnote b
%\item[$^{\mathrm{c}}$] This is footnote c
%\item[$^{\mathrm{d}}$] This is footnote d
%\end{list}
   \end{table}
%____________________________END    ___________________ One column table

In our models, we consider both internal and external heat sources.  
Externally
 we use the interstellar radiation field (ISRF) of Mathis, Mezger, 
$\&$ Panagia (1983) as the primary source of heating of the dust grains.
We have also considered the effects of an updated ISRF compiled by Evans 
(2001) for 
comparison.  This results in no significant qualitative difference 
in our results.
To study the effects of young stellar objects (YSOs), protostars, and other
embedded sources, we have also considered embedded sources with luminosities
ranging from $0 ~ L_\odot \leq L_\ast \leq 300 ~ L_\odot$.

The distribution of material in a star forming region is a topic of continuing
study.  
Both theoretical (e.g., Larson 1969; Shu 1977) 
and observational (e.g., Fuller \& Myers 1993; 
van der Tak et al. 1999; Evans et al. 2001) studies
suggest that the density distribution can be well fit by a power 
law.
On the other hand, recent work
(e.g., Henriksen, Andre, \& Bontemps 1997; Alves, Lada, \& Lada
2001; Whitworth \& Ward-Thompson 2001) predict/infer
more realistic density profiles,
including multiple power laws and Bonner-Ebert spheres.
Observations and detailed modeling by Evans et al. (2001) for a 
number of star-forming sources suggest that it may not always yet be
possible to observationally distinguish between power-laws and 
Bonner-Ebert spheres.  As a result, and as we wish to provide a 
simple yet somewhat realistic parameterization, we restrict our 
study here to power laws as given in equation (2).

Based upon models for cloud collapse and observations of star-forming
regions (e.g., Shu 1977; Whitworth \& Summers 1985; Foster \& Chevalier 1993;
Chandler \& Sargent 1993; Ward-Thompson et al. 1994; 
Andre et al. 1993), we model sources with density distribution 
indices in the range $0-2$, namely $m = 0$, $1$, and $2$. 
We arbitrarily take the outer shell radius to be
$r_\mathrm{out}=1$ pc, and the ratio of the inner and outer radii 
to be $r_0/r_\mathrm{out}
= 10^{-5}$.  This choice of inner radius is chosen to usually yield
dust temperatures of $\sim 1000\mathrm{K}$.  
We find that the choice of outer radius has little effect
on the resulting spectra. 

We take the grain opacity function
to be a power law of frequency as given in equation (3).
Laboratory experiments show that for 
crystalline grain material, $\beta \approx 2$ and for amorphous grain
material, $\beta \approx 1$ 
(see e.g., Tielens $\&$ Allamandola 1987; Henning, Michel, Stognienko 1995; 
Agladze et al. 1996;
Jager, Mutschke, \& Henning 1998; 
Fabian et al. 2001
).  Cox and Mezger
(1989) have shown that for $1 \leq \beta \leq 2$, the general behavior of the
opacity function should be covered in our models.  
Consequently, we take a 
value of $\beta = 1.5$.  We normalized our opacity function to Draine's (1985)
values at 2.2$\mu \mathrm{m}$ for the scattering component 
and at 190$\mu \mathrm{m}$ for
the absorption component.
These grain properties not only generally represent laboratory
data, but they also have the advantage that they 
reproduce observations of at least Orion
quite well (Doty \& Neufeld 1997).

Typical flux spectra for sources of varying luminosity and 
density distributions are shown in Fig. 1.  In Fig. 2 we show
typical temperature distribution for sources of varying luminosity and 
density distributions.

\begin{figure}
  \resizebox{\hsize}{!}{\includegraphics{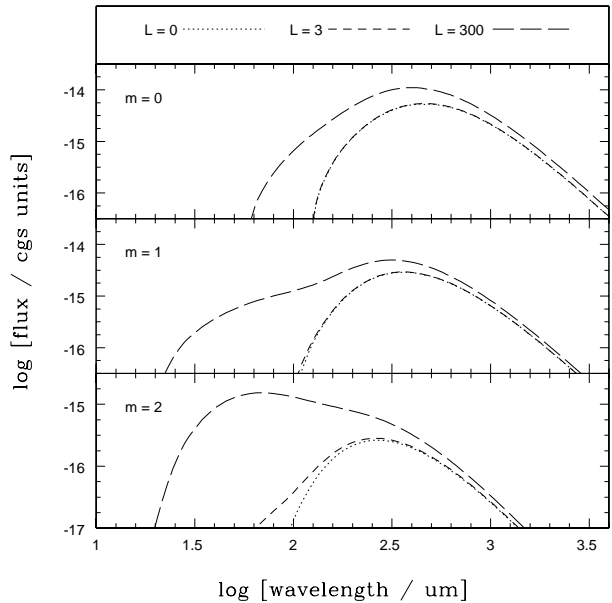}}
  \caption{Sample spectra for various models with $\tau_{0}=100$.  
  The three panels correspond to different values of the 
  density distribution index, $m$.  The different line
  types (dotted, short-dashed, long-dashed) correspond
  to various values of the central luminosity 
  (0, 3$L_{\odot}$, 10$L_{\odot}$ respectively).
  Note the effect of central luminosity on dust temperature,
  and hence on the short wavelength end of the spectra.}
  \label{flux}
\end{figure}

\begin{figure}
  \resizebox{\hsize}{!}{\includegraphics{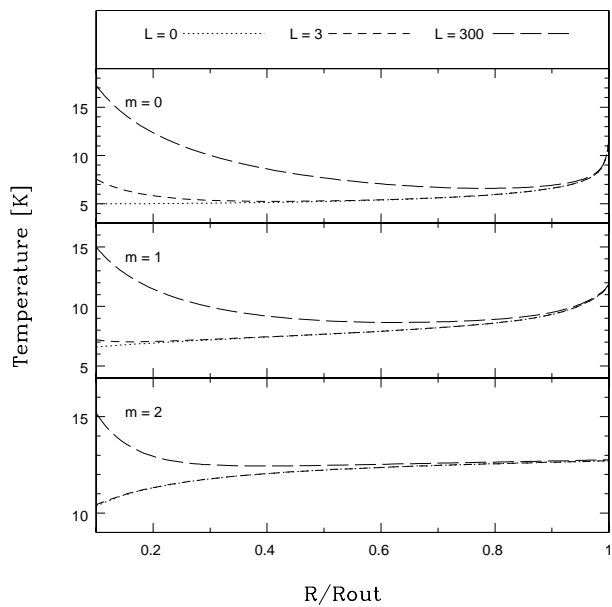}}
  \caption[]{Comparison of dust temperatures as a function of
  position in the source.  The models, lines, and panels are te same as in
  Fig. 1.  Notice how an increase in the internal luminosity
  leads to higher temperatures in the interior, while the 
  temperature on the outer edge is fixed by the ISRF.  Note also
  the correlation between similar temperature distributions here,
  and flux spectra in Fig. 1.}
  \label{temp}
\end{figure}

%==============================================================
\section{Dust Temperature}

Upon integrating equation (5) over all frequencies, it can be seen that
the thermal energy density varies as $T^{4+\beta}$.  As a result, a small 
change in temperature translates into a large change in the emergent spectrum.
Therefore, it is very important to determine the temperature distribution 
(Doty and Leung 1994; Shirley et al. 2000).  The effect of 
different temperature distributions can be seen in Figs. 1 and 2.  
The peak wavelength of emission as well as the width of the spectra are
affected by the dust temperature distribution.
  For the low luminosity 
embedded sources, the spectrum for the $m=2$ density distribution 
peaks at shorter wavelengths than the spectra for models with $m=1$ and
$m=0$.  This is consistent with the corresponding temperature
distributions in Fig. 2, where the $m=2$ models are generally warmer. 
On the other hand, for the low luminosity embedded sources, the
more uniform sources having $m=0$ and $m=1$ have a greater range of
temperatures, and in general have cooler dust.  The range in temperatures
results in a wider spectrum in Fig. 1.

This effect is also seen when comparing the models with $L_{*}=300L_{\odot}$
with those of lower luminosity.  In this case, the central source heats
the dust to 800-1000K on the inner edge, producing a very large range of
dust temperatures in the envelope.  The highest temperatures occur
with increasing optical depth and density distribution index, $m$, due to
radiation trapping in the interior.
Conversely, the inner dust 
temperature for a source
with either no or a low luminosty central source is between 5-15 K 
with the higher 
temperatures resulting from a lower optical depth and a higher density
distribution index, $m$.  These temperatures are not seen in Fig. 2 due 
to the axes chosen for the plot.
The corresponding emission spectra of the high luminosity cases 
both peak at a higher
flux and have emission over a broader range of wavelengths.

\subsection{Isothermal approximation}

As the intensity along a ray depends upon the temperature-dependent
emissivity integrated along the line of sight, it is difficult to determine
the temperature distribution in regions of star formation.  As a result,
it is easiest to assume the source is isothermal.  This assumption is 
best suited for sources which have either no embedded source, or only a very
low luminoity central source.  
As seen in Fig. 2, the temperature distribution is about constant 
for more than
$50$ per cent of the source for such cases.  
When the distribution is not constant, it only
changes by 2-5 K.  On the other hand, for higher luminosity central sources 
the range of temperatures
is 800-1000 K. 

In order to determine the appropriate isothermal temperature to assign to 
a given source, it is often easiest to consider the peak of the spectrum.  
We accomplish this using a Wiens Law modified to account for the wavelength 
dependance of the absorption coefficient, $Q_{\nu}$. 

In general, the peak of the emergent spectrum ($I_\nu$) for an opacity index 
$\beta$ can be given by 
\begin{equation}
T_\mathrm{peak}=\frac{f(\beta)}{\lambda_\mathrm{peak}
(\mu \mathrm{m})} \mathrm{K}.
\end{equation}
Here $f(\beta)$ is determined numerically, with results shown in 
Fig. 3.  As can be seen, the results are well-fit by
$f(\beta)=4620 e^{-0.2357\beta}$ with a correlation coefficient
of $r^{2}=0.9986$ for the range $1<\beta<2$.  In 
the case that $\beta=1.5$ (see Sect. 6 below), 
we find that $f(\beta)=3234$. 
This temperature can then be compared to other physical 
temperatures associated with
the source.  In this way it is possible to know what   
meaning, if any, can be assigned to this forced isothermal 
temperature.
 
\begin{figure}
  \resizebox{\hsize}{!}{\includegraphics{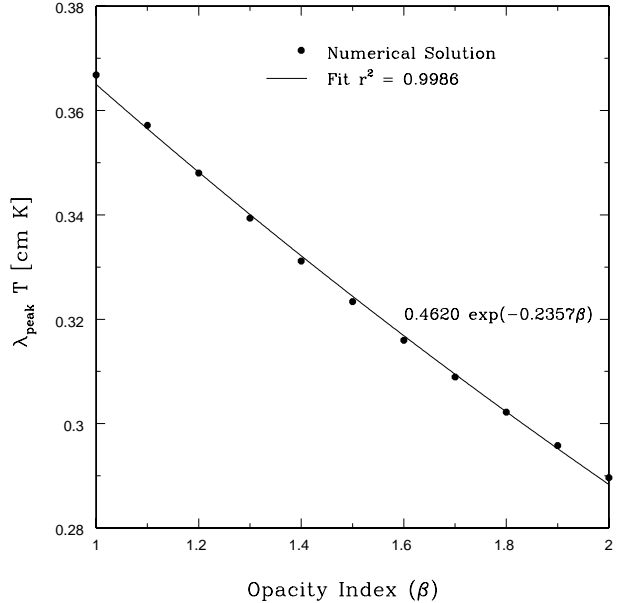}}
  \caption[]{The product of the peak wavelength and
  source temperature for a Wiens law modified to account
  for various opacity indicies.  The numerical results (data points) 
  are plotted as a function of the assumed opacity index. 
  The data are well-fit by the function (solid line) 
  $f(\beta)=4620e^{-0.2357\beta}$ with a correlation coefficient
  of $r^2 = 0.9986$.
  }
  \label{fbeta}
\end{figure}

One temperature used for comparison is the number-weighted grain average
temperature defined as
\begin{equation}
<T>_N = \frac{\int{T(r)n(r)4\pi r^2 dr}}{\int{n(r)4 \pi r^2 dr}}.
\end {equation}
This temperature is the average grain temperature.  Another 
comparison may be made 
with the energy-weighted temperature defined as
\begin{equation}
<T>_E = \left( \frac{\int{[T(r)]^{4+\beta}n(r)4 \pi r^2 ~ \mathrm{d}r}}
{\int{n(r)4 \pi r^2 ~\mathrm{d}r}}\right)^{1/(4+\beta)}.
\end{equation}
Physically, this temperature represents the average energy emitted by a grain.

In all models tested, except $m=2$, $L_{*}=300L_{\odot}$ as discussed below,
the peak temperature inferred
from equation (18) does not vary more than 0.5 K from
$<T>_E$. This can be understood as due to the fact that most of the energy
is radiated at a wavelength corresponding to the peak emission and so the 
corresponding temperature should be about the energy-weighted temperature.
 Also, the temperature agrees well with $<T>_N$. 
In Fig. 4, we plot $\Delta T \equiv T_\mathrm{peak}-<T>_N$ as a function of
the optical depth of the model for cases of different central luminosity.
The one outlying result is the $m=2$ case with a highly luminous embedded
source having $L_{*} = 300 L_{\odot}$.  In this case, the average 
difference between the inferred and number-weighted dust temperature is
about 50K.  
This difference can be understood by the fact that
in a centrally condensed source more dust will be affected by
a central heat source.  Consequently, the peak of the spectrum is shifted 
to shorter wavelengths, and is due to the 
central source rather than by the ISRF. 
As a result, the inferred temperature from the peak of the flux
spectrum is considerably too high and inapplicable to the problem at hand.
In all other cases in Fig. 4, the ISRF dominates the majority of 
the dust heating, and a few trends are immediately obvious. 
As optical depth increases, so does $\Delta T$, although
the largest difference is 1 K.  For $\tau_0 < 30$ the difference
is at most 0.6 K.  
In conclusion, deriving the dust temperature using
a modified Wiens Law will give a temperature that 
closely relates both the average
energy and the average grain temperature.

\begin{figure}
  \resizebox{\hsize}{!}{\includegraphics{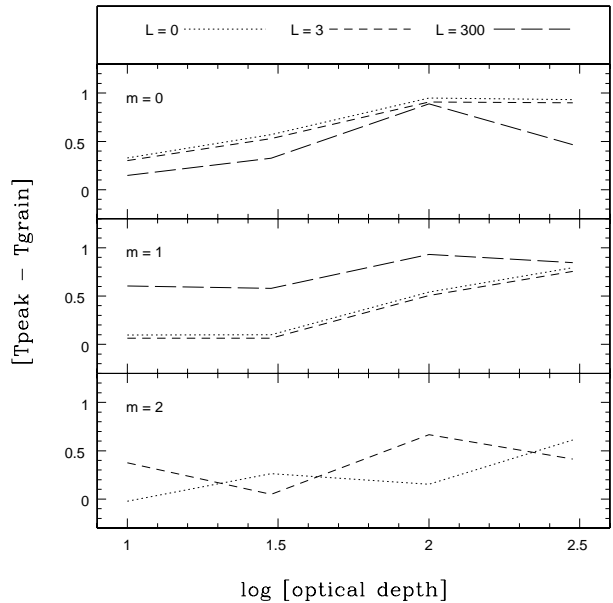}}
  \caption[]{Comparison of the temperature inferred 
  from the peak of the flux spectrum, with the number-weighted
  average dust temperature for the grid of models previously 
  discussed.  Note that the temperature inferred from the
  peak of the dust spectrum is a good measure of the 
  average (by number) dust temperature, often accurate to 
  within 0.5 K.}
  \label{dtemp}
\end{figure}
 
In order to see the effect of using an isothermal temperature 
on the emergent spectra,
in Fig. 5 we plot both the output spectra and a blackbody modified by the
absorption coefficient, normalized at the peak in the left panels. 
In the right 
panels we plot the fractional difference between 
the blackbody spectrum and the 
actual emergent spectrum.  
For the longer wavelength portion of the spectrum, the 
derived blackbody spectrum is at most about $20$ per cent in error, 
with larger discrepancies occuring 
at a smaller density distribution index, $m$.  
The $m = 2$ case is within $10$ per cent,
and the $m = 1$ case is within $15$ per cent.  
For wavelengths shorter than the peak, 
the errors tend to get worse.  In the $m=2$ case, the errors reach only 
about $5$ per cent, while for the $m = 1$ and $m=0$ cases, 
the errors can reach over $100$ per cent.   As the
luminosity of an embedded source increases, 
stellar emission makes it difficult to fit
the spectrum with a single temperature blackbody.
It should be noted that Fig. 5 shows the most favorable case 
for a single temperature fit,
since the lack of any internal heat source makes this the
most isothermal of all the models.  As a result
the other ($L_{*} \ne 0$) cases will be 
even more poorly fit by a single temperature. 

\begin{figure}
  \resizebox{\hsize}{!}{\includegraphics{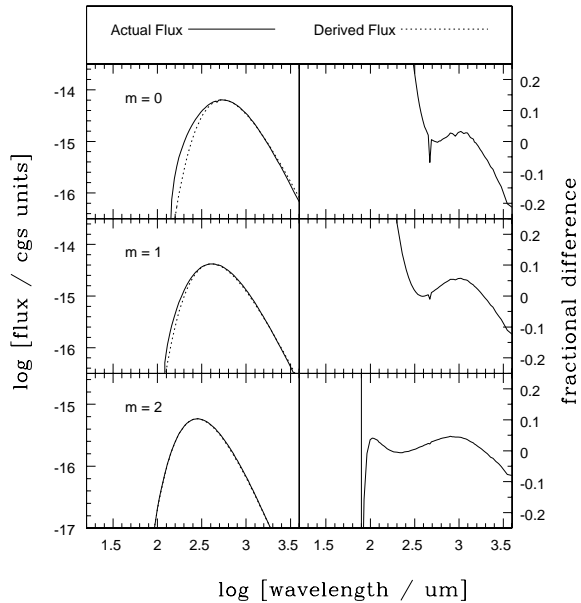}}
  \caption[]{The left hand panels show a comparison
  between the actual flux spectra (solid lines) and 
  dust-modified blackbody spectra at the temperatures
  inferred by the modified Wiens law in equation (18) 
  of the text (dashed lines).  Results shown here are for 
  $\tau_{0}=100$, and $L_{*}=0L_{\odot}$.
  Note the good agreement, especially near the peak.  
  For better comparison, the right hand panels show the fractional
  differences between the spectra.  Note that even though a single
  dust-modified blackbody fits well, the non-isothermal nature of 
  the sources can be clearly seen in the fractional differences.
  }
  \label{bbody.fracdiff}
\end{figure}

Although the peak of the output spectrum is a reasonable estimate for
either the number- or energy-weighted average dust temperature for
regions without strong internal sources, these single temperatures
do not represent the actual temperature distribution in the region.
As a result, isothermal fits do not generally duplicate the
output spectra to better than $10$ per cent.  Consequently, probes that
may be sensitive to the range of dust temperatures in a region
may require self-consistent radiative transfer modeling.

%==============================================================
\section{Dust Mass}

If the mass of a single grain is $m_g$, then the total dust mass is given by
\begin{equation}
M = m_g \int{n(r)4 \pi r^2 ~\mathrm{d}r}.
\end{equation}
Knowledge of the density 
distribution, $n(r)$, is necessary to determine the dust mass exactly, without
assumptions.  Although the form can be approximated
by a power law, its evaluation requires knowledge of both $r_0$ and $n_0$,
 which
are the inner radius and the amount of dust at the inner radius respectively.
These quantities are unknown, suggesting the use of more approximate
methods based upon dust emission.

\subsection{Dust mass from integrated luminosity}

The total integrated luminosity measures the total energy emitted per unit 
time and is given by
\begin{equation}
L = \int{L_\nu ~\mathrm{d}\nu}.
\end{equation}
Since dust grains can only emit what they absorb, the total energy emitted 
should be proportional to the optical depth, or dust mass, of the source.  A
relative mass can therefore be determined by taking the ratio of luminosities
of two different sources. The two basic underlying assumptions here is that
the sources are optically thin to the radiation of interest
and that the two sources being compared share the same temperature
distribution.  Only then will the dust mass be proportional
to the total emitted energy.
 We test this method by calculating the dust mass
for a series of models of increasing optical depth and different 
central luminosities. We normalize the models to the value at 
$\tau_0 = 10$.
In Fig. 6, we plot the normalized fractional difference in mass,
$\Theta$, defined as
\begin{equation}
\Theta \equiv \frac{M_\mathrm{derived} - M_\mathrm{actual}}{M_\mathrm{actual}}
\end{equation}
for varying optical depths.  The dust masses inferred by this method 
 are consistenly lower
than the actual dust masses for the same sources.
The median error is about $50$ per cent with errors
as great as $96$ per cent (i.e. a factor of $\sim 100$).  
The discrepancy increases
with the optical depth, and plateaus at higher optical depths
regardless of the luminosity.

\begin{figure}
  \resizebox{\hsize}{!}{\includegraphics{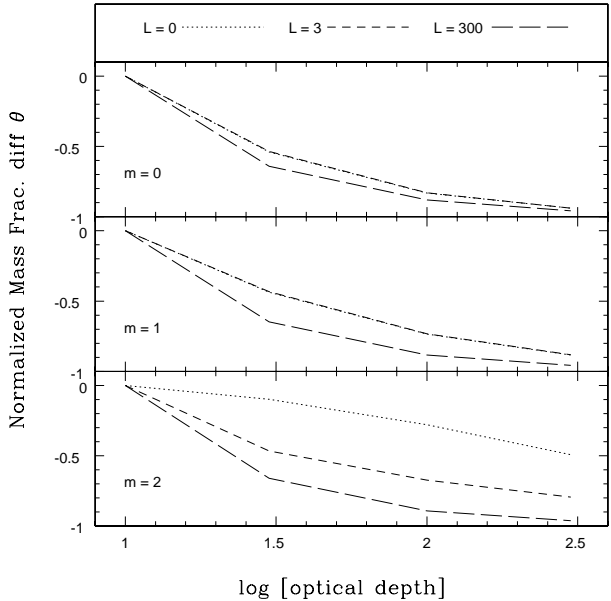}}
  \caption[]{The fractional difference between the dust mass inferred
  by considering the relative integrated luminosity between sources, 
  and the actual dust mass for the grid of models.  The results are
  normalized at the lowest optical depth.  Notice how the estimated dust
  masses are further in error with increasing optical depth and
  decreasing density contrast, due to the increasing opaquenss of
  the sources to radiation from the ISRF.}
  \label{nmass}
\end{figure}

The plateau for $\tau_0 > 100$ is due to the fact that the 
sources are no longer transparent to most IR radiation.  
As the optical depth increases
a greater percentage of incident energy from the ISRF is absorbed.
Eventually, at high enough optical depths, all
of the possible radiation will be absorbed leading 
to a constant total luminosity
and hence an inferred mass independent of $\tau$.
This effect is less dramatic in the $m = 2$ case because
most of the mass is centrally condensed, and therefore only radiation that 
penetrates to the center of the source will be absorbed. 

The consistently lower results can be understood in the following way.  The 
luminosity is proportional to the density distribution times the 
energy emitted per grain, which is the blackbody function modified by the 
absorption coefficient.  In order for the luminosity to be truly proportional
to the optical depth the energy emitted per grain needs to be the same for the
two sources being ratioed.  However, as the optical depth 
increases, the ability for radiation to heat the inside of the source 
attenuates as $e^{-\tau}$, and so the temperature distribution
between sources of different
optical depths are no longer going to be the same.  Therefore, the actual 
luminosity does not increase as quickly as the 
luminosities predicted using this
method and hence, the results are consistently lower.

For the $m=2$ case, the differences in the cases with different luminosities
can be explained as follows.  Most of the mass is located to the inside, and 
therefore most of the optical depth is located there as well.  Therefore, a 
majority of the source will be optically thin and so the same majority of the
source will be heated to a constant temperature.  This helps explain why the 
results for these cases are better than for the $m=1$ and $m=0$ cases.

\subsection{Dust mass from specific luminosity}

Hildebrand (1983) proposed that the dust mass of an optically thin and 
isothermal source can be related to the
observed flux of a given frequency and distance to the source via
equation (9).
In order to calculate the total dust mass, the mass of a grain, isothermal
temperature, and $Q_\nu$ need to be calculated or estimated. 
To alleviate
this problem, we take the ratio of dust mass for similar sources.  
In this way,
the only parameter that needs to be estimated is the temperature, 
which we calculate 
using the modified Wiens law in equation (18). 
We use the wavelength of peak emission as
the wavelength in equation (9).  
To evaluate this method we calculate the dust
mass for a series of models of increasing optical depths and different 
central luminosities, and compare them to the actual values.  We plot
the fractional difference as a function of optical depth in Fig. 7.

\begin{figure}
  \resizebox{\hsize}{!}{\includegraphics{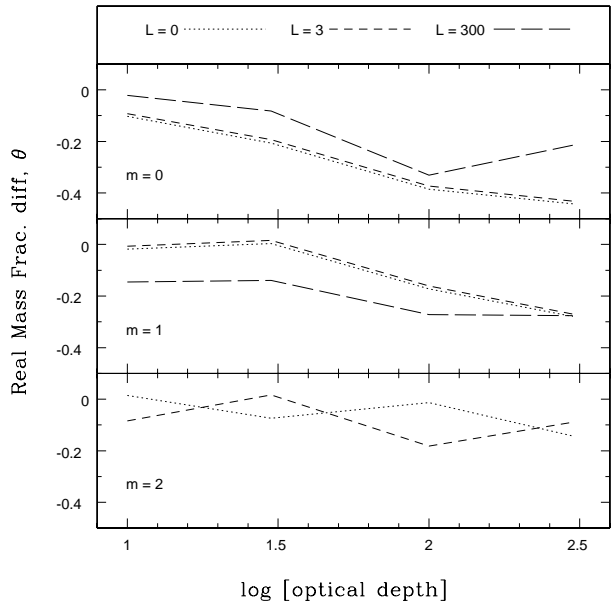}}
  \caption[]{The fractional difference between the mass inferred
  from the peak of the flux spectrum and the actual mass, for the
  grid of models discussed previously.  Notice the results are 
  generally better than the masses inferred by comparing 
  integrated luminosities.  This is related to the quality of the
  dust temperature from the peak as a measure of the number-weighted
  dust temperature.} 
  \label{rmass}
\end{figure}  

The results are quite good.  A median error of
about $25$ per cent is found for $m=0$, with errors between 
$10-20$ per cent depending on the central luminosity for $m=1$, 
and approximately $10$ per cent errors for $m=2$.  As the optical
depth increases, the differences also tend to increase.  As was the case for
calculating temperature, the mass inferred in the  $m=2$, $L=300$  
$~ L_\odot$ model is a factor of 50 times 
higher than the actual mass.  The error is due to the inability to determine
accurately an average temperature of the source.  

In most cases, as the luminosity increases, the error increases as well.  
However, in the case of $m=1$, $L=300 ~L_\odot$ we find that the errors 
are less than 
without an embedded star.  With a constant density distribution, a larger
proportion of the mass is found to the outside of the source 
and so there is a 
larger change
in temperature to the outside of the source (see Fig. 2).
However, when there is an embedded star with a high luminosity, 
the rest of the
source is heated and it causes a smaller change in temperature
to the outside of the source,
giving the appearance of a more isothermal source (see Figs 2 and 4).

\section{Opacity Function}

The grain opacity index, $\beta$, defines the form of the opacity function.  
The effects are also evident in the flux spectrum.  A lower $\beta$ has the 
effect of increasing the amount of emission at wavelengths 
longer than the visible
range relative to the emission in the visible.  
Therefore, a lower optical index
will broaden the spectrum.  Conversely, a higher $\beta$ has the effect of 
narrowing the spectrum.

\subsection{Finding $\beta$ as a ratio of luminosities}
A more common, and potentially easier method for determining 
$\beta$ is to use the ratios of 
luminosities from two different wavelengths, as shown in equation (12). 
 This method relies on two
assumptions: that the RJA holds, and that the source is optically thin.  
Again, we have evaluated this technique for models of increasing optical
depths and different luminosities.  Fig. 8 
 shows the opacity index derived between
a reference wavelength, denoted by the various symbols, and a variety of 
longer wavelengths.  Our results are consistently lower that the actual value
but asymptotically approach it for longer wavelengths.  The source is still 
optically thin over the wavelength range we are looking 
(at $650\mu \mathrm{m}$, $\tau=0.002$) 
and so the deviations
are due to the RJA.  The presence of a temperature distribution has the effect
of adding up multiple blackbodies.  The RJA will only hold true if we are on
the long wavelength end of a blackbody corresponding to the coolest dust.  For
the coolest dust in our models ($\sim 5$ K), 
the shortest wavelength that satisfies
the RJA is $\sim 2800 \mu \mathrm{m}$.

\begin{figure}
  \resizebox{\hsize}{!}{\includegraphics{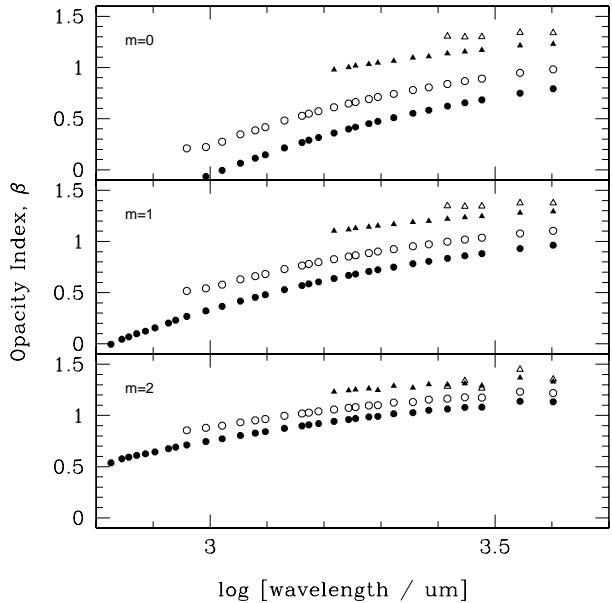}}
  \caption[]{The inferred dust opacity index for models with 
  $\tau_{0}=300$.  The results for other optical depths are 
  qualitatively similar.  The opacity indicies are derived from the slope of
  the FIR spectrum, and plotted versus the shortest wavelength of the
  wavelength-pair used in the slope determination.  Like-symbols
  correspond to variations in the longer wavelength of the pair.
  Note that even for wavelengths at which the source is transparent,
  one must go to extremely long wavelengths to ensure that the 
  RJA holds.}
  \label{opacity}
\end{figure}

As can be clearly seen in Fig. 8, there is some wavelength short of
which the inferred opacity indicies are inapplicable.  
We call this wavelength the cutoff wavelength, and define it to be
the shortest wavelength in the pair used in equation (12)  
for which $\beta$ is determined to within $20$ per cent.  In Fig. 9, we plot
$\lambda_{\mathrm{cutoff}}$ as a function of optical depth
for the models considered.  We see that in general, wavelengths 
greater than $1000-2000 \mu\mathrm{m}$ should be used in determining
$\beta$ from the slope of the FIR spectrum,
so long as the dust continuum dominates the free-free emission --
a significant problem in some sources
.  These results can be
considered a means of inferring the validity of the RJA.

\begin{figure}
  \resizebox{\hsize}{!}{\includegraphics{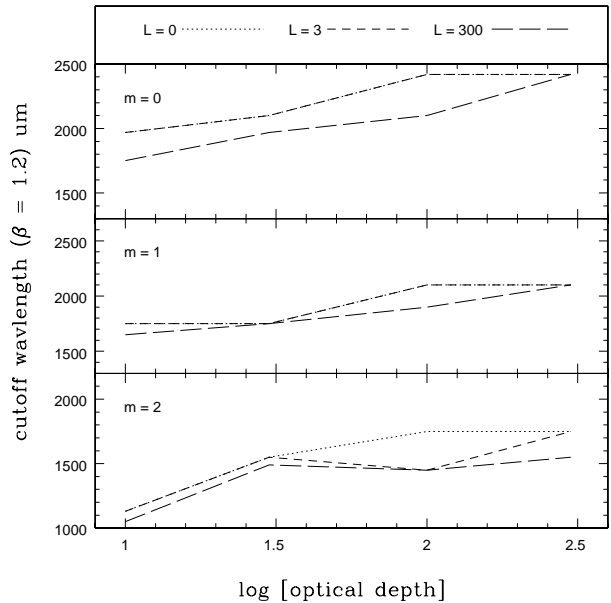}}
  \caption[]{The variation in the cutoff wavelength with 
  optical depth for the models considered.  As discussed in the text,
  the cutoff wavelength is the shortest wavelength for which the 
  slope of the FIR will yield an opacity index correct to within
  20 per cent}  
  \label{cutoff}
\end{figure}

\subsection{Fitting the spectrum to determine $\beta$}
The opacity index may also be determined by treating 
it as a fitting parameter while
fitting the spectrum.
It is easiest and somewhat common 
(see e.g., McCarthy, Forrest, \& Houck 1978; Sopka et al. 1985)
to assume an isothermal source in making these fits.
While this makes the fitting easier, it also removes the effect that the 
temperature distribution has on the width of the spectrum.   
Consequently, the derived opacity indices are 
artificially raised to compensate.
To try and alleviate this problem, we have only fit the longer wavelength 
end of the spectrum as it should be more optically
thin, as well as sample the cooler (and hence more isothermal) dust.  
This can be seen in Fig. 1
 where we see that a star does not have a significant effect on the presence
 of the spectrum at longer wavelengths.

Based upon this reasoning, and with the advent of recent 
and upcoming far IR/submm
observational data (e.g. SCUBA, SIRTF, Herschel, etc.) we fit the wavelength 
range $500 \mu \mathrm{m} < \lambda < 900 \mu \mathrm{m}$ 
for models of different density
distributions and increasing optical depth shown in Table 1.  
Our results are shown
in Fig. 10.  The errors in this method range from $0$ to $60$ per cent.
In the case of $m = 1$ and $2$, the derived indices are too high, 
with slightly better results as the optical depth increases.
For the $m=0$ case, the derived indices
are too high by about $7$ per cent for optical depths $\tau_0 < 30$, and at an
optical depth of $\tau_0 = 300$ the derived indices are too low by $15$
per cent.

\begin{figure}
  \resizebox{\hsize}{!}{\includegraphics{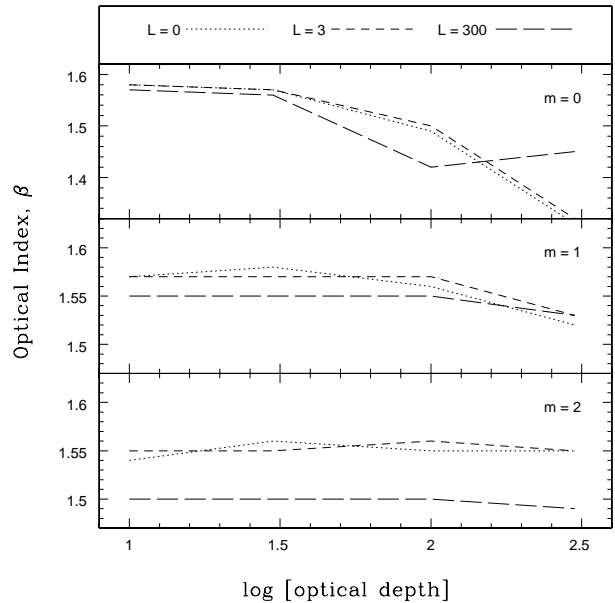}}
  \caption[]{The dust opacity index inferred by fitting the 
  dust spectrum for the grid of models considered.  The results 
  are much better than those inferred from the slope of the FIR spectrum,
  and are often good to better than 10 per cent.} 
  \label{opacityfit}
\end{figure}

For luminosities $L \leq 3  L_\odot$, 
the derived opacities follow the same trend as the starless
core models.  But for luminosities $L > 3  L_\odot$, 
the derived opacity indices tend to
be lower, and more accurate by an average of $50$ per cent.  
Even at $500 \mu \mathrm{m}$, the
star affects the spectrum noticeably (see Fig. 1).  The spectrum is 
artificially broadened by the star, and thus the derived opacities are 
lowered.  In effect, the errors caused by the isothermal assumption are 
 counteracted by the extra luminosity from an embedded star.

In order to account for the non-isothermal nature of the sources,
we also fit the spectrum with two blackbodies.  
This simulates a two-temperature 'core-halo'
 model, with a warm exterior shell heated by the ISRF, and a cool 
interior shell.  These fits reduced our error to a maximum of $7$ per cent.
However, the five fitting parameters make it 
more difficult to obtain reasonable results due to the presence of
local minima, with the accuracy of the fitting depending 
more significantly on the initial guess
for the parameters.  In this case, the fits have a maximum error of 
0.1 in $\beta$, with most having errors of 0.05 or less, 
representing a median error of $
\sim 3$ per cent.

We also considered the use of different wavelength ranges on the derived
opacity indices.  We did not test the use of shorter 
wavelengths, due to contamination 
of the spectrum by the embedded sources.
Using the range from
$500\mu \mathrm{m}$ to the end of the spectrum 
(about $3000\mu \mathrm{m}$), the derived indices are 
uniformly low for a single blackbody fit, with 
an average $\beta = 1.2$.  This 
effect is due to the addition of emission from the coolest dust, which  
causes the spectrum to broaden past $900\mu \mathrm{m}$.  
Therefore, even when longer
wavelengths are chosen, the results are not appreciably better, 
as multiple tempertures are still sampled.

\section{Density Distribution}

The density distribution is crucial to our understanding of the processes and 
regions of star formation.  
Both observations and theory indicate that 
typical distributions can be well-described by
a power law.  As a result we assume
a power law distribution in all of our models.  The density index, $m$, 
helps to determine the emergent radiation of a source.  A centrally 
condensed source ($m = 2$) is less affected by the ISRF and more affected by
embedded sources.  In contrast, a constant distribution ($m=0$) is affected by
the ISRF more than by an embedded star (see Fig. 1).

\subsection{Density distribution in a resolved source}
Yun and Clemens (1991) extended a note by Tomita et al. (1979) 
regarding the relationship of the column density power index to the
density distribution power index. 
Using equation (17), a column density map can be formed from the
density distribution index.
By matching the derived and calculated column density profiles
$m$ can be constrained. 

To test the reliability of this method we have calculated the density 
distribution index for a series of models with 
different density distributions, optical depths,
and central luminosities.
We consider the wavelengths of 
$175 \mu \mathrm{m}$ and $215 \mu \mathrm{m}$, as they
should be near the peak of emission.
As the effects of beamsize are instrument-dependent, and may
potentially confuse the analysis, we consider an infinitely
small pencil beam for most of this work in order to keep the
possible sources of uncertainty to a minimum.

Based upon the difficulty in determining an 'edge' to the source,
and based upon our experience in these fits, we find that 
choosing impact parameters in the range $0.03~r_{\mathrm{out}} < p
< 0.3~r_{\mathrm{out}}$ provide the optimal fits over a wide range
of source parameters.
We also tested different impact parameters, both closer to the center, and out
towards the edge.  We found that both of these cases gave poor results.  
As the beam probes closer to the edge, part of the beam is off the source.  
As a result
there is an artificial lowering in the calculation of the column density.  As 
the beam probes toward the center, the contamination of the spectrum by
the embedded source yields poor results.

In Fig. 11, 
we plot $\Delta m \equiv m_\mathrm{calculated} - m_\mathrm{actual}$ 
for the conditions discussed above. 
We find that the results are moderate at best.  We find  
a median $|\Delta m|$ of $0.1 - 0.4$, which 
corresponds to a difference from as 
low as $5$ per cent to over $100$ per cent.
The maximum deviations range between 0.5 and 0.8, corresponding
to fractional differences between $25$ per cent to over $100$ per cent.

\begin{figure}
  \resizebox{\hsize}{!}{\includegraphics{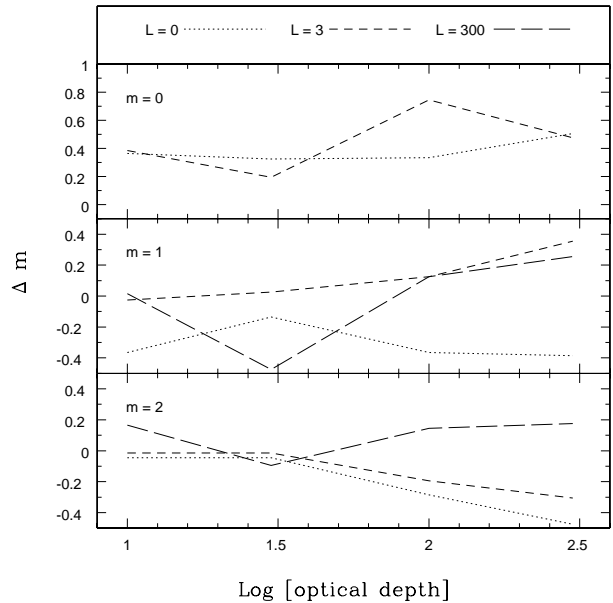}}
  \caption[]{The difference between the inferred and actual dust
  density distribution for the grid of models considered.  Notice
  that the results are better for more centrally-condensed sources.} 
  \label{mdiff}
\end{figure}

In the $m = 2$ case, we see that the results are good for the 
low luminosities and optical depths.
This is due to the fact that the source is fairly 
isothermal in these cases.  As the optical depth increases,
the inferred density distribution index also tends to worsen.
This is because at higher optical depths, the temperature 
gradient is larger.  The problem is not due to the optical depth because
the source is transparent at these wavelengths.
For the high luminosity
case, we find that the values are not as good as the low luminosity 
cases at low
optical depths, but tend to be slightly better as optical depth increases.  We
find that generally $\Delta m > 0$, meaning that the
inferred index is greater than the actual index.  This is because we see a 
slight increase in temperature towards the center and so we get more emission
at smaller radii.  As a result, we infer more mass towards the center, giving
us a larger distribution index.

We find that the $m = 1$ case is harder to understand.  We find that for no
embedded star that there is a monatomic decrease in temperature.  As a result
we have more inferred mass towards the outside, 
artificially lowering the index.  

In the $m=0$ case, the results are uniformly too 
high with a median difference 
of 0.4.  The high luminosity case
did not produce any results.

For comparison, we have also tested the effects of using longer
wavelengths of 
$670 \mu \mathrm{m}$ and $800 \mu \mathrm{m}$, for comparison with
general wavelength ranges probed by SCUBA and other recent observations.
We find that using these wavelengths results in somewhat worse results.
In particular, the resulting mean differences between the
density distribution indicies inferred at these longer wavelengths
and those inferred near the peak of the spectrum are $6$, $15$, 
and $20$ per cent for $m=0, 1, 2$ respectively.  In all cases, the indicies
inferred near the peak of the flux spectrum provide more accurate
results.    

We also tested the effects of different beam sizes by varying the
beam size from 15 arcsec to 4 arcmin (roughly from SCUBA to IRAS beam sizes),
corresponding to roughly 60 and 4 beams across the source respectively at an 
assumed distance of 500 pc.  
We find that the effect of beam size is relatively small for
smaller beams.  In particular, for beams approximately appropriate
for SOFIA ($\sim 6$ arcsec) and SCUBA ($\sim 15$ arcsec), the 
inferred density distribution has a median variation from the
infinitesimal beamsize of only $\sim 5$ per cent.  Larger beams, such as
$\sim 4'$ similar to IRAS produce results with at least a $70$ per cent
difference.  In both cases, these results are due to the ability 
to resolve the source, and the effect of the source only partially
filling the beam for pointings near the edge.  As a result, it appears
that for regions with small temperature gradients, relatively 
well-resolved observations provide results that are
rather insensitive to the beam size.

\subsection{Detailed modeling of the flux spectrum 
to find the density distribution}
Another approach to try and help determine the 
density distribution is to model
the flux spectrum.  
In order to test this method we produced artificial 
observations for a base model similar to that described in 
Sect. 3, but having $m=1.5$ to simulate a more 'intermediate' value.  
In order to more closely resemble actual observations,
we randomly added noise to the simulated spectra.
The noise amplitude was as large as $25$ per cent 
of the simulated observations,
with a median noise of $13$ per cent over the entire spectrum.
We then ran over 2000 models that bracketed this base model in two 
parameters, optical depth and density distribution index.  
Based upon the results of Sect. 6, we assume that the grain properties
could be determined relatively accurately, and thus treat $\beta$ as
known.

We evaluated the models by calculating
the chi-squared difference between the flux spectrum 
of the test models with the simulated observations.
In Fig. 12,  we plot the $99.999$ 
per cent confidence interval for this result.  
We can see that for any reasonably constrained optical depth, 
the density distribution index, $m$, is constrained to the range 
$1.2 \leq m \leq 1.8$, which is a difference of $20$ per cent from the 
value of $m=1.5$ adopted in the model.
However, if the optical depth can be determined to within
$10$ per cent, 
the error in determining the density index, 
$m$, becomes $\sim 7$ per cent.

\begin{figure}
  \resizebox{\hsize}{!}{\includegraphics{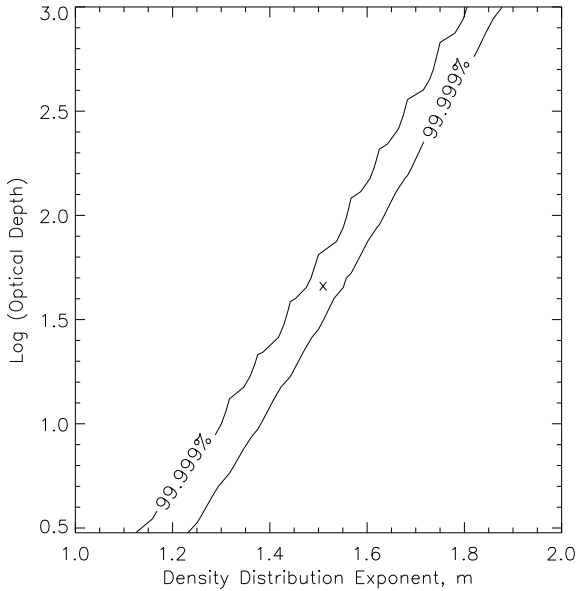}}
   \caption[]{A plot of the 99.999 per cent confidence contour in the
   determination of the density distribution exponent as a function
   of the optical depth of the source.  While there is a definite
   line of degeneracy, uncertainties in the optical depth of an 
   order of magnitude yield uncertainties in the density distribution
   exponent of only 0.25 -- or about 15 per cent.}
\end{figure}
 
Uncertainties of a similar magnitude in $m$ (i.e. $\Delta m \sim 0.2 - 0.3$)
can also result from uncertainties in the ISRF 
(Young \& Evans 2002), the source geometry (Doty et al. 2002b), 
and the assumed value of $\beta$.  
However, it may be possible to constrain these parameters by
other data (e.g., bolometric luminosity for 
pre-protostellar cores, source morphology, and the results
of Sect. 6).  
It is interesting to note that
even when these parameters are not well constrained, the
uncertainties in $m$ are on the order of the uncertainties
implied by the optical depth alone, and smaller otherwise.  

\subsection{Surface brightness profile}
Finally, as the smaller beamsizes of recent and upcoming instruments
make it increasingly possible to spatially resolve star-forming
regions (see e.g., Evans et al. 2001) it is interesting to 
consider the implications of the expected surface brightness across the
source.  Shirley et al. (2000) presented models of the normalized
intensity profile as a function of impact parameter for 
semi-analytic cases similar to equation (17) with various assumptions
regarding the temperature profile in the dust envelope.  They found
that the assumptions of isothermality and the RJA could be important.

For comparison, in Fig. 13 we plot the expected surface brightness
at $345 \mu \mathrm{m}$
as a function of impact parameter for $m=0,1,2$, and 
$L_{*}=0,3,300 L_{\odot}$.
This wavelength is chosen as it should be near the peak of the
source spectrum, and is observable by SCUBA.
These results are based upon detailed modeling as
decribed earlier.  In this case the dust temperature distribution
is determined self-consistently, and the RJA is not applied.

\begin{figure}
  \resizebox{\hsize}{!}{\includegraphics{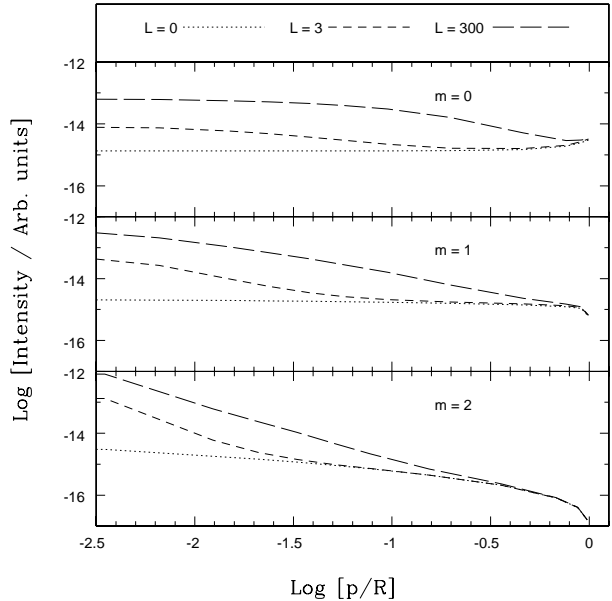}}
   \caption[]{The surface brightness for a pencil beam 
   at $345 \mu \mathrm{m}$
   across the model sources as a function of impact parameter.
   All results shown are for $\tau_{0}=100$, with only 
   qualtiative differences for other optical depths.   
   The different line types correspond to varying central 
   luminosities, and the panels to different density distribution
   exponents.  Note the effects that both the central source
   and density distribution have on the surface brightness, 
   suggesting the need for detailed modeling to best interpret
   such observations.
   } 
\end{figure}

From these results, some conclusions may be drawn.  First, for well-
resolved sources, it appears that it is possible to infer the dust
density distribution index, $m$.  This is especially true if the
source luminosity is somewhat constrained, and more importantly
if good signal-to-noise data exist near the source edge without
contamination from beam dilution, etc.  Second, it may be possible
to infer the existence of an internal heat source from the surface
brightness distribution.  This is a problem which is the subject of
an upcoming paper (Doty \& Moore 2002).  Finally, these results
show an apparently wider variation in surface brightness across the
source than those from approximate methods (see e.g., Fig. 13 of Shirley
et al. 2000).  The differences are due to the discrepancies 
between the assumed
dust temperature distribution in the approximate methods, 
and the self-consistent distribution determined through the modeling.
This underscores the importance of detailed modeling of emergent
radiation for these and any other but the most simple 
star-forming regions.

\section{Conclusions}
We have critically evaluated current semi-analytical methods in analysing
regions of low-mass star-formation.  
In particular, we have investigated the validity of the underlying
assumptions for the methods and determined the conditions under which 
these methods are reliable.  Based upon this work, we find that:

1.  The temperature distribution plays a significant role in determining the 
spectrum of a source.  As there are no current semi-analytic methods for 
determining the temperature distribution explicitly, 
a common assumption is that
these sources are isothermal.  
Although an isothermal temperature with some physical basis can be found 
using a modified Wiens Law (Fig. 3 and equation [18]), 
using this temperature does not accurately determine the other source 
parameters (See Sections 4, 5, and 6).

2.  It is difficult to reliably determine 
the total dust mass without an accurate 
estimate of grain properties, density distribution, 
and the distance to the source
(Section 5).  However, a ratio of mass from 
two similar sources can be determined using
the observed flux at a given frequency.  
The dust masses determined by this method are 
accurate to within $50$ per cent for all cases where 
the estimated temperature is in reasonable 
agreement (5 K) with the average grain temperature -- namely
for regions with no or weak internal sources.  
It is also possible to determine
a ratio of dust masses using the integrated luminosity.  
This method is only applicable for low optical depths,
as errors approach $100$ per cent for $\tau_{0} \ge 100$.

3.  For an optically thin source and in a region 
of frequencies where the Rayleigh Jeans approximation holds,
the opacity index, $\beta$, can be determined using 
the ratio of two different frequencies as in equation (12).  
In order to determine the opacity index to 
within $20$ per cent for all optical depths, the pair of wavelengths used
need to be longer than about $2500\mu \mathrm{m}$, due to the 
failure of the RJA and the lack of isothermality of the source.
This may be problematic due to the existence of free-free 
emission which may dominate the thermal dust continuum for
some sources.
Another method for determining 
the opacity index is by fitting the spectrum.  
With a single blackbody fit, $\beta$ can be
determined to within $60$ per cent for all cases.  
If, instead, a two temperature fit is used,
the results for $\beta$ are improved considerably, 
with a maximum error of only $7$ per cent.

4.  The density distribution index, $m$, 
can be determined for a resolved source that is 
isothermal along any line of sight by using the ratio 
of intensities at two different
wavelengths.  For a theoretically small pencil beam, the results have
an error of up to $100$ per cent.  
Errors increase as the beam size gets larger, growing dramatically
for pointings in which the source does not fill the beam.  
The errors also increase somewhat as the wavelengths used decrease.  
Finally, due to the isothermal assumption,
it is important to probe impact parameters that are 
neither too close to the edge of the source nor too close to the core.  
We find $0.03 r_{\mathrm{out}} < p < 0.3 r_{\mathrm{out}}$ works well
in the cases studied here.

5.  It should be possible to infer the dust density distribution,
$m$, from the surface brightness distribution across the source.
This, however, requires good signal-to-noise data toward the
source edge.  The range in surface brightness distributions 
is due to the differences
in the actual temperature distributions in the sources, highlighting
the need for self-consistent modeling to ensure proper interpretation
of observational data.

%
%                                                One column figure
%------------------------------------------------- new & old rates 
%   \begin{figure}
%      \resizebox{\hsize}{!}{\includegraphics{fig3.eps}}
%%      \vspace{5cm}
%      \caption[]{Destruction rates per molecules of CO$_{2}$ by H$_{2}$  
%		 ($\frac{\mathrm{d} n(\mathrm{CO}_2)}{\mathrm{d}t} /
%		 n(\mathrm{CO}_2)$ s$^{-1}$) for the GL and MFW rates.  
%		 The labels give $\log[n(\mathrm{H}_2)]$. 
%		 The solid lines show the ranges over which the
%		 rates are quoted, and the dashed lines 
%		 their extrapolations.
%              }
%         \label{Fignewoldrates}
%   \end{figure}
%%
%%______________________________________________________________

\section*{Acknowledgements}
      We are grateful to the referee, 
      Derek Ward-Thompson, for helpful suggestions, and 
      to Neal Evans for discussions and useful comments.  
      We thank Ewine van Dishoeck for hospitality at
      Leiden University where the work was initially begun. 
      This work was partially supported under
      a grant from The Research Corporation, a bezoekerbeurs from
      the Netherlands Organization for Scientific Research (NWO), and
      supplementary support from the Battelle Corporation 
      through Denison University.

\label{lastpage}


\begin{thebibliography}{}

   \bibitem[2000]{Abrahametal2000} Abraham, P., et al. 2000,
      A\&A, 354, 965

   \bibitem[1996]{Agladzeetal1996} Agladze, N. I., et al. 1996, 
      ApJ, 462, 1026
      
   \bibitem[2001]{alvesetal2001} Alves, J. F., Lada, C. J., 
      \& Lada, E. A. 2001, Nature, 409, 159      
      
   \bibitem[1996]{AndreWardThompsonMolte1996} Andre, P., 
      Ward-Thompson, D., \& Molte, F. 1996, A\&A, 314, 625
      
   \bibitem[1993]{AndreWardThompsonBarsony1993} Andre, P.,
      Ward-Thompson, D., \& Barsony, M. 1993, ApJ, 406, 122  
      
   \bibitem[2000]{AndreWardThompsonBarsony2000} Andre, P., 
      Ward-Thompson, D., \& Barsony, M. 2000, in: Protostars and
      Planets IV, eds. V. Mannings, A. Boss, \& S. Russell
      (Tucson: Univ. of Arizona Press), 59       

   \bibitem[1991]{butneretal1991} Butner, H. M., Evans II, N. J.,
      Lester, D. F., Levreault, R. M., \& Strom, S. E. 1991,
      ApJ, 376, 636 
      
   \bibitem[1993]{ChandlerSargent1993} Chandler, C. J., \&
      Sargent, A. I. 1993, ApJ, 414, 129   
      
   \bibitem[1993]{Churchwell1993} Churchwell, E. B. 1993, in:
      ASP Conf. Ser. 35, Massive Stars, Their Lives in the
      Intestellar Medium, eds. J. P. Cassinelli \& E. B.
      Churchwell, ASP, 35
      
   \bibitem[1999]{Churchwell1999} Churchwell, E. B. 1999, in:
      The Physics of Star Formation and Early Stellar Evolution II,
      eds. C. J. Lada \& N. D. Kylafis, Kluwer, 515         
      
   \bibitem[1989]{CoxMezger1989} Cox, P., \& Mezger, P. G., 
      1989, ARA\&A, 1, 49     

   \bibitem[1994]{DotyLeung1994} Doty, S. D., \& Leung, C. M. 1994,
      ApJ, 424, 729 
      
   \bibitem[1997]{DotyNeufeld1997} Doty, S. D., \& Neufeld, D. A. 1997,
      ApJ, 489, 122
      
   \bibitem[2002]{Dotyetal2002a} Doty, S. D., van Dishoeck, E. F.,
      van der Tak, F. F. S., \& Boonman, A. M. S. 2002a, A\&A, in press 
      
   \bibitem[2002]{Dotyetal2002b} Doty, S. D., et al. 2002b, in preparation 
  
   \bibitem[2002]{DotyMoore2002} Doty, S. D., \& Moore, M. M. 2002,
      in preparation 
   
   \bibitem[1985]{Draine1985} Draine, B. T. 1985, ApJS, 57, 587 

   \bibitem[1988]{EganLeungSpagna1988} Egan, M. P., Leung, C. M.,
      \& Spagna, G. F., Jr. 1988, Comput. Phys. Comm., 
      48, 271
      
   \bibitem[2001]{EvansRawlingsShirleyMundy2001} Evans II, N. J.,
      Rawlings, J. M. C., Shirley, Y. L., \& Mundy, L. G. 2001, 
      ApJ, 557, 193
      
   \bibitem[2001]{Fabianetal2001} Fabian, D., et al. 2001, 
      A\&A, 378, 228         
      
   \bibitem[1993]{FosterChevalier1993} Foster, P. N., \& 
      Chevalier, R. A. 1993, ApJ, 416, 363
      
   \bibitem[1993]{FullerMyers1993} Fuller, G. A., \& Myers, P. C.
      1993, ApJ, 418, 273      

   \bibitem[2001]{Gradyetal2001} Grady, C. A., et al. 2001, AJ,
      122, 3396 
      
   \bibitem[1989]{Helou1989} Helou, G. 1989, in IAU
      Symp. 135, Interstellar Dust, ed. L. J. Allamandola \&
      A. G. G. M. Tielens (Dordrecht: Reidel), 285 

   \bibitem[1997]{Henrikseneta1997} Henriksen, R., 
      Andre, P., \& Bontemps, S. 1997, A\&A, 323, 549

   \bibitem[2000]{Henningetal2000} Henning, Th., 
      et al. 2000, A\&A, 364, 613
      
   \bibitem[1995]{HenningMichelStognienko1995} Henning, Th., 
      Michel, B., \& Stognienko, R. 1995, Planet. Space Sci., 43, 1333
      
   \bibitem[1983]{Hildebrand1983} Hildebrand, R. H. 1983,
      QJRAS, 24, 267        

   \bibitem[1999]{DUSTY} Ivezic, Z., Nenkova, M., \& Elitzur, M. 1999, 
      User Manual for DUSTY, Univ. Kentucky Internal Rep.

   \bibitem[1998]{JagerMutschkeHenning1998} Jager, C., Mutschke, H.,
      \& Henning, Th. 1998, A\&A, 332, 291
      
   \bibitem[2001]{KertonMartinJohnstoneBallantyne2001} Kerton, C. R.,
      Martin, P. G., Johnstone, D., \& Ballantyne, D. R. 2001,
      ApJ, 552, 601   
      
   \bibitem[1998]{KrugelSiebenmorgenZotaChini1998} Kr\"ugel, E.,
      Siebenmorgen, R., Zota, V., \& Chini, R. 1998, A\&A, 331, L9      

   \bibitem[1969]{Larson1969} Larson, R. B. 1969, MNRAS, 145, 271

   \bibitem[1996]{Launhardtetal1996} Launhardt, R., et al. 1996,
      A\&A, 312, 569
      
   \bibitem[1997]{LaunhardtWardThompsonHenning1997} Launhardt, R.,
      Ward-Thompson, D., \& Henning, Th. 1997, MNRAS, 288, L45      

   \bibitem[1978]{MFH1978} McCarthy, J. F., Forrest, W. J., 
      \& Houck, J. R. 1978, ApJ, 224, 109

   \bibitem[1997]{MH1997} Men'shcikov, A. B., \& Henning,
      Th. 1997, A\&A, 318, 879
      
   \bibitem[1976]{SchmidBergkScholz1976} Schmid-Bergk, J., \&
      Scholz, M. 1976, A\&A, 51, 209      
      
   \bibitem[2000]{Shirleyetal2000} Shirley, Y. L., Evans II, N. J., 
      Rawlings, J. C., \& Gregersen, E. M. 2000, ApJS, 131, 249   

   \bibitem[1977]{Shu1977} Shu, F. H. 1977, ApJ, 214, 488
   
   \bibitem[1987]{ShuAdamsLizano1987} Shu, F. H., Adams, F. C., 
      \& Lizano, S. 1987, ARA\&A, 25, 23
      
   \bibitem[1999]{SiebenmorgenKrugelChini1999} Siebenmorgen, R., 
      Kr\"ugel, E., \& Chini, R. 1999, A\&A, 351, 495      

   \bibitem[1985]{Sopkaetal1985} Sopka, R. J., et al. 1985, 
      ApJ, 294, 242
      
   \bibitem[1987]{TielensAllamandola1987} Tielens, A. G. G. M., \&
      Allamandola, L. J. 1987, in Interstellar Processes, ed. 
      D. J. Hollenbach \& H. A. Thronson, Jr. (Dordrecht: Reidel), 397      

   \bibitem[1999]{vdt99} van der Tak, F. F. S., et al. 1999,
      ApJ, 522, 991
      
   \bibitem[2000]{vdt2000} van der Tak, F. F. S., et al. 2000, 
      ApJ, 537, 283      

   \bibitem[1998]{visseretal1998} Visser, A. E., Richer, J. S., 
      Chandler, C. J., \& Padman, R. 1998, MNRAS, 301, 585
      
   \bibitem[1999]{VoshchinnikovKrugel1999} Voshchinnikov, N. V.,
      \& Kr\"ugel, E. 1999, A\&A, 352, 508      
      
   \bibitem[1994]{WardThompsonScottHillsAndre1994} Ward-Thompson, D.,
      Scott, P. F., Hills, R. E., \& Andre, P. 1994, MNRAS, 268, 276   

   \bibitem[2000]{WardThompsonZylkaMezgerSievers2000} Ward-Thompson, D.,
      Zylka, R., Mezger, P. G., \& Sievers, A. W. 2000, A\&A, 355, 1122
      
   \bibitem[1985]{WhitworthSummers1985} Whitworth, A., \& Summers,
      D. 1985, MNRAS, 214, 1   
      
   \bibitem[2001]{Whitworthwardthompson2001} Whitworth, A. P., 
      \& Ward-Thompson, D. 2001, ApJ, 547, 317      

   \bibitem[2002]{YoungEvans2002} Young, C., \& Evans, N. J. 
      2002, ApJ, submitted      


\end{thebibliography}
\end{document}